\documentclass[aps,prl,twocolumn,superscriptaddress,numerical]{revtex4-2}
\usepackage[utf8]{inputenc}
\usepackage[english]{babel}
\usepackage[T1]{fontenc}
\usepackage{amsmath,amssymb,dsfont}
\usepackage[colorlinks,
citecolor=purple,
linkcolor=purple,
urlcolor=purple]{hyperref}
\usepackage{tikz}
\usepackage{lipsum}
\usepackage{dsfont}
\usepackage{physics}
\usepackage{float}
\usepackage{braket}
\usepackage{orcidlink}
\usepackage{multirow}
\usepackage{soul}

\graphicspath{{images}}

\newcommand{\Haar}{{\text{Haar}}} 
\newcommand{\Cl}{{\text{Cl}}}         
\newcommand{\Wg}{{\text{Wg}}} 
\newcommand{\spur}{\Xi} 

\renewcommand{\section}[1]{\textbf{\emph{#1}}.\!}
\renewcommand{\subsection}[1]{{\emph{#1}}.---}

\begin{document}

\title{Quantum State Designs via Magic Teleportation}

\author{Hugo Lóio~\orcidlink{0000-0002-4011-8180}}
\affiliation{Laboratoire de Physique Th\'eorique et Mod\'elisation, CNRS UMR 8089, CY Cergy Paris Universit\'e, 95302 Cergy-Pontoise Cedex, France}

\author{Guglielmo Lami~\orcidlink{0000-0002-1778-7263}}
\affiliation{Laboratoire de Physique Th\'eorique et Mod\'elisation, CNRS UMR 8089, CY Cergy Paris Universit\'e, 95302 Cergy-Pontoise Cedex, France}

\author{Lorenzo Leone~\orcidlink{0000-0002-0334-7419}}
\affiliation{Dipartimento di Ingegneria Industriale, Università degli Studi di Salerno, Via Giovanni Paolo II, 132, 84084 Fisciano (SA), Italy}
\affiliation{Dahlem Center for Complex Quantum Systems, Freie Universit\"at Berlin, 14195 Berlin, Germany}

\author{Max McGinley~\orcidlink{0000-0003-3122-2207}}
\affiliation{TCM Group, Cavendish Laboratory, University of Cambridge, Cambridge CB3 0HE, UK}

\author{Xhek Turkeshi~\orcidlink{0000-0003-1093-3771}}
\affiliation{Institut für Theoretische Physik, Universität zu Köln, Zülpicher Straße 77, D-50937 Köln, Germany}

\author{Jacopo De Nardis~\orcidlink{0000-0001-7877-0329}}
\affiliation{Laboratoire de Physique Th\'eorique et Mod\'elisation, CNRS UMR 8089, CY Cergy Paris Universit\'e, 95302 Cergy-Pontoise Cedex, France}

\begin{abstract}
We investigate how non-stabilizer resources enable the emergence of quantum state designs within the projected ensemble. Starting from initial states with finite magic and applying resource-free Clifford circuits to scramble them, we analyze the ensemble generated by performing projective Pauli measurements on a subsystem of the final state.
Using both analytical arguments and large-scale numerics, we show that the projected ensemble converges towards a state $k$-design with an error that decays exponentially with the $k$-th Stabilizer Rényi Entropy of the pre-measurement state, via a Magic-Induced Design Ansatz (MIDA) that we introduce. We identify a universal scaling form, valid across different classes of magic initial states, and corroborate it through numerical simulations and analytical calculations of the frame potential.  For finite-depth Clifford unitaries, we show that the timescales at which state designs emerge are controlled by the transport of magic. We identify a ``magic teleportation'' mechanism whereby non-Clifford resources injected locally spread through Clifford scrambling and measurements across distances beyond the lightcone. Our results demonstrate how a small and controlled amount of magic suffices to generate highly random states, providing a systematic route toward generating quantum state designs in early fault-tolerant devices.
\end{abstract}

\maketitle

\section{Introduction}
Quantum computers promise transformative computational power that would enable the solution of various classically intractable problems~\cite{Feynman1982Simulating,Lloyd1996UniversalSimulators,Shor1997Algorithms,Grover1997Search,BernsteinVazirani1997QCT,AaronsonArkhipov2011LinearOptics,Zhong2020PhotonicAdvantage,Kretschmer2025}. 
To reach this goal, implementations must move beyond the present-day noisy intermediate-scale quantum regime and achieve fault tolerance~\cite{Preskill2018NISQ,Quantinuum2025,Aasen2025,Peham2025}. 
This challenge is met by quantum error correction, which is based on Clifford operations and stabilizer codes~\cite{shor1995,gottesman1997stabilizercodesquantumerror,gottesman1998theory,Nielsen_Chuang_2010,Bluvstein2023,Gidney2021stimfaststabilizer}. 
While the most straightforward operations to implement fault-tolerantly are typically Clifford gates, which preserve this stabilizer structure, these resources alone are insufficient to outperform classical computers~\cite{Aaronson2004,bravyi2016improved,Bravyi2019simulationofquantum}. 
Magic, or non-stabilizer, resources are required to unlock the full  power of quantum computation~\cite{chitambar2019quantum,liu2022manybody,Leone_2022,Leone_2024,tirrito2024quantifying,Turkeshi_2023,turkeshi2025pauli,Turkeshi_2024_2,jasser2025stabilizerentropyentanglementcomplexity,viscardi2025interplayentanglementstructuresstabilizer,Iannotti2025entanglement,cusumano2025nonstabilizernessviolationschshinequalities,bittel2025operationalinterpretationstabilizerentropy,varikuti2025impactcliffordoperationsnonstabilizing,tirrito2025anticoncentrationnonstabilizernessspreadingergodic,zhang2025stabilizerrenyientropytransition,qian2025quantum,moca2025nonstabilizernessgenerationmultiparticlequantum,dowling2025magic,bera2025nonstabilizernesssachdevyekitaevmodel,masotllima2024stabilizer,aditya2025mpembaeffectsquantumcomplexity,hernándezyanes2025nonstabilizernessquantumenhancedmetrologicalprotocols,falcão2025magicdynamicsmanybodylocalized,sticlet2025nonstabilizernessopenxxzspin,tirrito2025universalspreadingnonstabilizernessquantum,zhang2024quantummagicdynamicsrandom,cao2025gravitationalbackreactionmagical,Tarabunga_2024,qiant25,qian2025,hang2025,qiant2024a,frau2025,fan2025disentangling,huang2024cliffordcircuitsaugmentedmatrix,ding2025evaluating,korbany2025longrangenonstabilizernessphasesmatter,tarabunga2025efficientmutualmagicmagic,szombathy2025independentstabilizerrenyientropy,hou2025stabilizerentanglementenhancesmagic,hoshino2025stabilizerrenyientropyconformal,tarabunga2024magictransitionmeasurementonlycircuits,tirrito2025magicphasetransitionsmonitored,wang2025magictransitionmonitoredfree,santra2025complexitytransitionschaoticquantum,Haug2025probingquantum,Haug2023stabilizerentropies,Haug_2023_1, Lami_2023_2, Lami_2024,tarabunga2023manybody,tarabunga2024nonstabilizerness,du2025certifyinglocalizablequantumproperties,monaco2025nonstabilizernesscostquantumstate}. 
Crucially, schemes to prepare magic states with an arbitrarily small error exist that are compatible with fault-tolerant, or logical, Clifford circuits \cite{bravyi2005universal,desilva2025cliffordhierarchyqubitqudit,Campbell2014,Campbell2017,Gorman2017,Silva2024,Zihan2025,Niroula2024,Turkeshi2024,PaetznickReichardt2013Transversal,Litinski2019MSCNotCostly}. This suggests that doped, or magic-augmented, Clifford circuits are particularly well-suited to near-term fault-tolerant devices~\cite{Haferkamp2022,zhou2020single,Leone2021quantumchaosis,magni2025quantumcomplexitychaosmanyqudit,Leone2024LearningTDoped,Chia2024SingleCopyTDoped,Gu2024DopedStabilizerManyBody,mao2025qudit,szombathy2025spectralpropertiesversusmagic,magni2025anticoncentration,leone2025noncliffordcostrandomunitaries,Zhang2025DesignsMagicAugClifford,liu2025classicalsimulabilitycliffordtcircuits,nakhl2025stabilizer,fux2025disentanglingunitarydynamicsclassically,Vairogs2025,scocco2025risefallnonstabilizernessrandom,Paviglianiti_2024, Bejan_2024_1, Fux_2024}. 

\begin{figure*}
\begin{minipage}[c]{0.60\textwidth}
\includegraphics[width=\columnwidth]{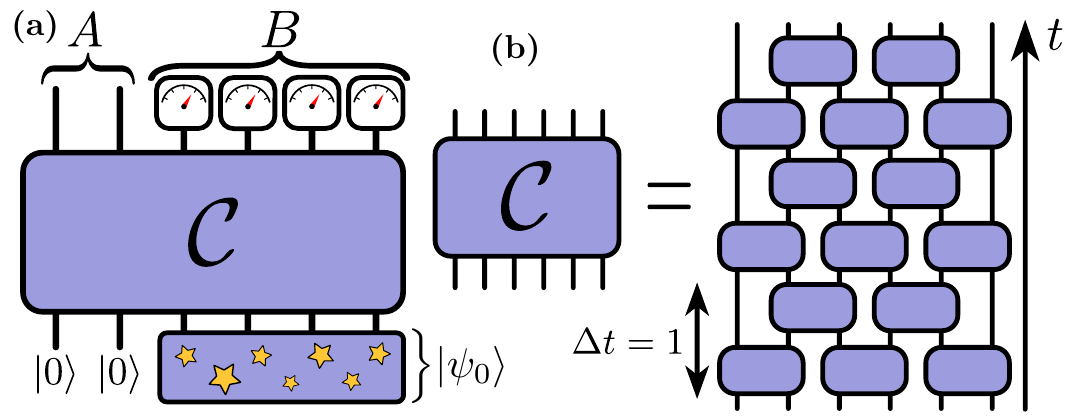}
\end{minipage}
\hspace{-4mm}
\begin{minipage}[c]{0.39\textwidth}
\resizebox{!}{0.33\textwidth}{
\renewcommand{\arraystretch}{1.5}
\setlength{\tabcolsep}{4pt}  
\begin{tabular}{|c|c|}
    \hline
     \multicolumn{2}{|c|}{Deep circuits ($L_A \gg 1$, $k>1$)} \\
     \hline
     No magic &  
     $\displaystyle \Delta^{(k)}_A = \frac{1}{k!} \prod_{i = 1}^{k-1} ( 2^{L_A} + i) - 1 \equiv \mu$ \\
    \hline
     One $T$-gate &  
     $\Delta^{(k)}_A = \mu \left( 1/2 + 2^{-k} \right)$ \\
    \hline
     Finite magic $M_k$ &
     $\Delta_A^{(k)} = \mu \exp\left[ (1-k) M_k \right] $  \\
    \hline \hline
    \multicolumn{2}{|c|}{Shallow circuits} \\
    \hline
    Local gates &  $\tau = O(\log L)$\\
    \hline
    Long-range ($\alpha < 2$) & $\tau = O(1)$ \\
    \hline
    Long-range ($\alpha > 2$) & $\tau = O(\log L)$ \\
    \hline
\end{tabular}
}
\end{minipage}

    \caption{
    Scheme of the main protocol used in this work. 
    (a) A Clifford operator $\mathcal{C}$ is applied onto an initial state $\ket{\psi_0}$ with known finite magic injected somewhere in partition $B$.
    The projected ensemble in $A$ is obtained by measuring out the degrees of freedom of subsystem $B$ in the $Z$ basis.
    The vertical lines represent qubit degrees of freedom.
    (b) The Clifford operator $\mathcal{C}$ will be mainly decomposed in a brickwork circuit of random 2-qubit Clifford gates with open boundary conditions and depth (time) $t$.
    On the right, a table that synthesises the main findings of the work. 
    $\Delta_A^{(k)}$ quantifies the distance between the $k$-th moments of the Haar ensemble and the projected ensemble, $M_k$ is the $k$-th stabilizer Rényi entropy, $\tau$ is the convergence timescale to the time-asymptotic (deep circuit) result, and $\alpha$ is an interaction parameter for a long-range circuit (different from the one in (b)) that suppresses the interaction range as it increases.
    For deep circuits, $\Delta_A^{(1)} = 0$ for any initial state (the projected ensembles of Clifford scrambled states are always at least 1-designs).
    The shallow circuits section assumes the magic is injected at a distance $O(L)$ from partition $A$.
    }
    \label{fig_protocol_scheme}
\end{figure*}

A particularly important task that relies on magic resources is the generation of random states that resemble the Haar ensemble \cite{Arute2019Supremacy,Wu2021StrongAdvantage, Zhu2022RCS60Qubits,Morvan2024RCSPhaseTransitions,Gao2025Zuchongzhi3,DeCross2025RCSGeometries,Madsen2022PhotonicAdvantage}.
Random states have many applications in quantum information, 
from tomography to demonstrations of quantum advantage
\cite{Arute2019Supremacy,Wu2021StrongAdvantage, Zhu2022RCS60Qubits,Morvan2024RCSPhaseTransitions,Gao2025Zuchongzhi3,DeCross2025RCSGeometries,Madsen2022PhotonicAdvantage}
and cryptography
\cite{ji2018pseudorandom, aaronson2023quantumpseudoentanglement},
and are also central to our understanding of quantum many-body  physics
\cite{Page1993AverageEntropy, hayden2007black, nahum2017quantum, Nahum2018}.
In particular, these applications require the formation of $k$-designs---ensembles that reproduce the first $k$ moments of Haar-random states~\cite{knill1995approximationquantumcircuits,Gross2007,Harrow2009,Gross_2016,Brandao2016LocalDesigns,grevink2025glueshortdepthdesignsunitary,west2025nogotheoremssublineardepthgroup,cui2025unitarydesignsnearlyoptimal,Haferkamp2022randomquantum,zhu2016cliffordgroupfailsgracefully,Brand_o_2016,hunterjones2019unitarydesignsstatisticalmechanics,Schuster2025,laracuente2024approximateunitarykdesignsshallow,lami2025quantum,lami2025anticoncentration,sauliere2025universalityanticoncentrationchaoticquantum,dowling2025freeindependenceunitarydesign,haferkamp,silvia,fritzsch2025freeprobabilityminimalquantum,schuster2025strongrandomunitariesfast,Varikuti2024unravelingemergence}. 
While state designs can be generated by sampling random unitary gates, one can also leverage the measurement-induced randomness inherent to quantum mechanics.
This approach is naturally advantageous for certain platforms, such as analog quantum simulators \cite{McGinley2023ShadowTomography,RevModPhys.80.1061}.
By performing projective measurements on a subsystem of a \emph{single} deterministically scrambled state, one generates an ensemble of post-measurement states in the remaining subsystem, a construction known as the \textit{projected ensemble} \cite{Cotler2021EmergentDesigns,HoChoi2022ExactDesigns,Lami2025}.
Projected ensembles are essential for our understanding of deep thermalization and the underlying nature of quantum chaos \cite{Ippoliti2022DeepThermalization,Mark2024MaxEntDeepThermalization,fava2025designs,Bhore2023DeepThermalizationConstrained, Lucas2023GeneralizedDeepThermalization, Chang2025DeepThermalizationU1,Lin2023ProbingSignStructureMIE}.
Furthermore, recent work has shown that measurements can accelerate the emergence of designs by concentrating, or ``teleporting,'' randomness across the system.
However, measurement-induced randomness remains largely unexplored in the physically relevant setting of Clifford circuits doped by limited amounts of magic.

This work addresses this problem by studying the formation of $k$-designs in the projected ensembles of Clifford circuits doped with a finite amount of magic, as illustrated in Fig.~\ref{fig_protocol_scheme}.
An initial magic state $\ket{\psi_0}$ on $L$ qubits is prepared and evolved under a Clifford circuit $|\psi_t\rangle=\mathcal{C}_t|\psi_0\rangle$, whose geometry depends on the model considered.
Each layer consists of random two-qubit Clifford gates $U_{x,y}\sim\mathfrak{C}_2$, drawn from the group generated by the Hadamard $H=(X+Z)/\sqrt{2}$,  phase $S=\sqrt{Z}$, and $\mathrm{CNOT}=|0\rangle\langle 0|\otimes I+|1\rangle\langle 1|\otimes X$ gates.
These gates leave the magic of the global state invariant, $M_k(|\psi_0\rangle)=M_k(|\psi_t\rangle)$, which we quantify through the Stabilizer Rényi Entropies~\cite{Leone_2022,Leone_2024} $M_k(|\psi\rangle) = (1-k)^{-1}\log \sum_{s\in \mathcal{P}_L} \frac{1}{2^L}\langle \psi|s|\psi\rangle^{2k}$, where $\mathcal{P}_L=\lbrace I,X,Y,Z\rbrace^{\otimes L}$ are the unsigned Pauli strings.
Stabilizer Rényi Entropies can be efficiently measured in experiments by joint measurements with $O(k)$ copies of the state ~\cite{Oliviero_2022,haug2024efficient}.
When qubits in a subsystem $B$ are projectively measured in a Pauli basis, the state of the remaining qubits $A$ is disturbed. This process, which can in principle alter the amount of magic in the system, generates a random ensemble of post-measurement states 
$\mathcal{E}=\lbrace |\psi_A(z_B)\rangle \, , p(z_B)\rbrace$, where $z_B$ labels the possible outcomes, which occur with Born probabilities $p(z_B)=\mathrm{tr}_A[\langle \psi_t|z_B\rangle\langle z_B|\psi_t\rangle]$, and $|\psi_A(z_B)\rangle=\langle z_B|\psi_t\rangle/\sqrt{p(z_B)}$ the post-measurement states. 

The central questions we address are: (i) \emph{How much initial magic is required for the projected ensemble $\mathcal{E}$ to approximate a $k$-design?} (ii) \emph{How do circuit locality and depth control the timescales of randomness generation, particularly in architectures of current and near-term relevance?}
To quantify the convergence to $k$-designs, we use the $k$-th frame potential of the projected ensemble 
\begin{equation}\label{eq_frame_potential}
    F_\mathcal{E}^{(k)}  = \sum_{z_B, z_B'} p(z_B) p(z_B') \left| \langle \psi_A(z_B) | \psi_A(z_B') \rangle \right|^{2k} 
  \, .
\end{equation}
The relative deviation $\Delta_A^{(k)} := (F_{\mathcal{E}}^{(k)}/F_H^{(k)})-1$ from the Haar-random value $F_H^{(k)}= \binom{2^{L_A}+k-1}{k}^{-1}$ can be used to upper bound the distance to a $k$-design. Specifically, $\mathcal{E}$ forms an $\epsilon$-approximate $k$-design (with trace distance error) whenever $\Delta_A^{(k)}<\epsilon^2$~\cite{Mok2025}.

Our main results, supported by analytical arguments and extensive numerical simulations,  are summarized in Fig.~\ref{fig_protocol_scheme}. 
In particular, for sufficiently deep Clifford circuits, our numerical results demonstrate that the projected ensemble converges to a quantum state design \emph{exponentially fast in the amount of magic resources of the initial state}, quantified by the k-stabilizer entropy $M_k(|\psi_0\rangle)$.
This behavior is fully consistent with our analytical bounds, which show that the deviation $\Delta_A^{(k)}$ decreases at a rate that outpaces the growth of magic in the initial state, as measured by the second stabilizer entropy $M_2$.
Finally, we analyze an alternative protocol in which magic is instead injected through measurements and find, quite remarkably, that it leads to exactly the same behavior.
For shallow Clifford circuits, the convergence of the frame potential depends strongly on gate locality: it saturates at depth $O(\log L)$ in local architectures, while for long-range connectivity with decay exponent $\alpha < 2$, convergence is achieved at constant depth, $O(1)$.
We should comment that the problem of magic-induced randomness in the projective ensemble has directly received attention recently, see for example ~\cite{Vairogs2025,Zihan2025}. Here we give a first analytical and full numerical treatment valid for any $L_A $ and $L$.

\begin{figure*}[t]
    \centering
    \includegraphics[trim={5 0 10 0}, clip, width=\linewidth]{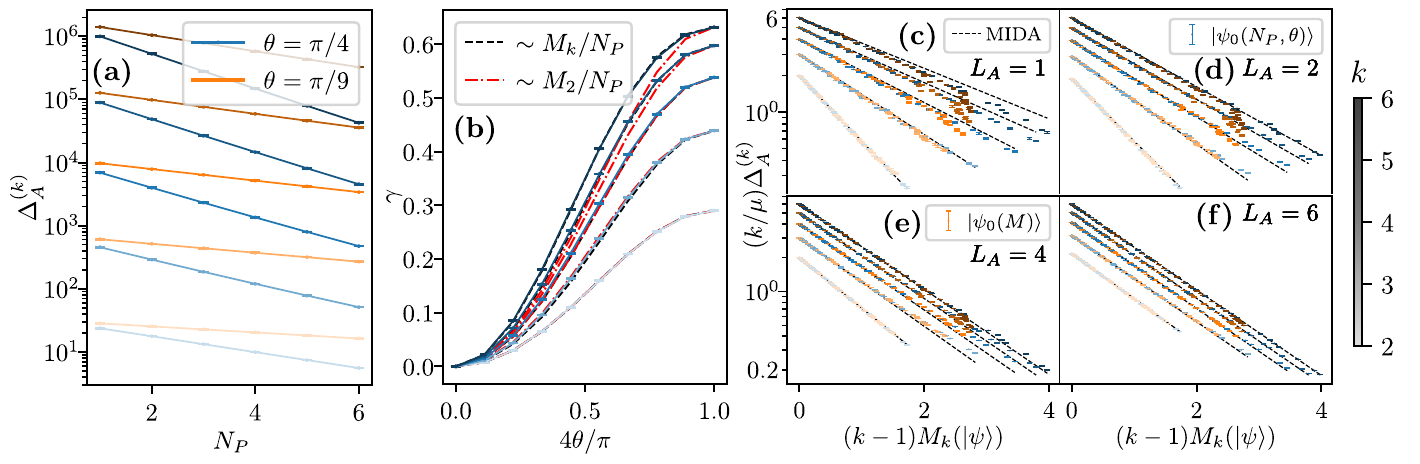}

    \vspace{-5pt}
    \caption{
    (a) Projected ensemble normalized distance to Haar $\Delta_A^{(k)}$, as a function of the initial number of magic qubits $N_P$ with phase $\theta \in \{\frac{\pi}{4}, \frac{\pi}{9}\}$. Different values of $k$ are explored (see colorbar on the right). (b) Decay rate of the Haar distance with the number of initial magic qubits, i.e.\ $\Delta_A^{(k)} \sim \exp(-\gamma N_P)$. We plot $\gamma$ as a function of the phase $\theta$, for different $k$. Dashed lines represent the $k$-th Stabilizer Rényi Entropy density of the pre-measurement state $M_k(\ket{\psi})/N_P$, up to a normalization constant. Dash-dotted lines represent the $2$-Stabilizer Rényi Entropy density, $M_2(\ket{\psi})/N_P$, which clearly fails to capture the behavior of the data. (c,d,e,f) Projected ensemble normalized distance to Haar $\Delta^{(k)}_A$ as a function of $(k-1)M_k(\ket{\psi})$. For visualization purposes, the $y$-axis is rescaled by $k/\mu$ (see Eq.~\eqref{eq_mu}). The blue curves correspond to initial product phase states with $N_P \in \{1,\dots, 6\}$ and $\theta \in [0, \pi/4]$. The orange curves correspond to random initial states with fixed magic $M \in [0, 2]$ (protocol in~\cite{SMAT}). The system size is $L = 22$ in all plots, while $L_A = 1$ in (c), $L_A = 2$ in (d), $L_A = 4$ in (e), and $L_A = 6$ in (a,b,f).
    }
    \label{fig_deep}
\end{figure*}

\section{MIDA for deep Clifford circuits}
Let us first consider the asymptotic limit of deep Clifford circuits, where $C$ is a global random Clifford transformation acting on $L$ qubits. 
In this limit, we uncover a direct connection between the design order of the projected ensemble and the magic of the initial state: the distance $\Delta_A^{(k)}$ decays exponentially with the stabilizer entropy $M_k(|\psi_0\rangle)$.

To corroborate this statement, we begin by presenting two analytically tractable regimes. 
First, we consider a system with no magic resources, fixing $\ket{\psi_0}$ without loss of generality. 
After application of a global random Clifford, the pre-measurement state $|\psi_\infty\rangle \equiv \mathcal{C} |\psi_0\rangle$ becomes a random stabilizer state. 
Owing to the properties of stabilizer measurements, all post-measurement states $\ket{\psi_A(z_B)}$ share the same stabilizer generators up to signs (see e.g.~Appendix 9D in Ref.~\cite{Bejan2025} or \cite{SMAT}). 
Consequently, the overlap between any two states $\ket{\psi_A(z_B)}$ and $\ket{\psi_A(z_B’)}$ in $\mathcal{E}$ satisfies $\left|\langle\psi_A(z_B)|\psi_A(z_B’)\rangle\right|\in\{0,1\}$. Substituting into Eq.~\eqref{eq_frame_potential} then immediately yields $F_\mathcal{E}^{(k)}=F_\mathcal{E}^{(1)}$ for all $k$.
Let us note that the frame potential for $k = 1$ corresponds to the purity of the reduced density matrix in $A$, namely $F_\mathcal{E}^{(1)} = \Tr \left[ \rho_A^2 \right]$, where $\rho_A = \mathrm{tr}_B[\mathcal{C}|\psi_0\rangle\langle \psi_0|\mathcal{C}^\dagger]$. 
Since this expression involves two copies of the Clifford unitary, the average over Cliffords coincides with the Haar average, yielding
\begin{equation}\label{eq_F1}
    \mathbb{E}_{\mathcal{C} \sim \mathfrak{C}}[F_\mathcal{E}^{(1)} ]=\frac{2^{L-L_A}+2^{L_A}}{2^L+1}\;,
\end{equation}
where $L_A=|A|$ and $\mathbb{E}_{\mathcal{C} \sim \mathfrak{C}}[\bullet]$ denotes the Clifford average. Asymptotically, for any initial state, $\lim_{L\to\infty} \mathbb{E}_{\mathcal{C} \sim \mathfrak{C}}\left[F_\mathcal{E}^{(1)}\right]=2^{-L_A}$. 
Consequently, in the absence of magic resources, the Haar distance of the projected ensemble is 
\begin{equation}\label{eq_mu}
    \mu \equiv  \Delta^{(k)}_A = \frac{1}{k!}\prod_{i = 1}^{k-1} \left( 2^{L_A} + i\right) - 1 \, \, \, \text{for} \, \, \,  k > 1 \, ,
\end{equation}
while $\Delta^{(1)}_A = 0$.
Thus, for stabilizer states the projected ensemble is restricted to a 1-design~\cite{Bejan2025, SMAT}.
It is interesting to contrast this result with the global state $|\psi_\infty\rangle=\mathcal{C}|\psi_0\rangle$, which instead forms a 3-design on the full system~\cite{Gross_2016}. In essence, projective measurements deteriorate this property, lowering the design order to one. 

We now analyze the case where the system is doped with the minimal amount of magic resources: a single $T$-state $|\psi_0\rangle = |0\rangle^{\otimes L-1} \otimes |T\rangle$, where $|T\rangle=(|0\rangle + e^{i \pi/4}|1\rangle)/\sqrt{2}$. 
In this setting, the evolved state can be written as a superposition of two stabilizer states that differ by the sign of one of their stabilizer generators. 
After measuring subsystem $B$, this structure restricts the possible overlaps between post-measurement states in $A$. A detailed analysis~\cite{SMAT}, shows that these overlaps can only take the values in $\{0,1,1/\sqrt{2}\}$. This restricted set of possibilities allows us to write the frame potential of the projected ensemble in the simple form
\begin{equation}\label{eq_avg_frame_1Tmeas}
\mathbb{E}_{\mathcal{C} \sim \mathfrak{C}}\left[F_\mathcal{E}^{(k)} \right] = 2^{-L_A} + C_2 \left[\left(\frac{1}{2}\right)^k - \frac{1}{2}\right],
\end{equation}
where $C_2$ is a constant that depends on the subsystem size $L_A$ and can be determined from the second frame potential $F_\mathcal{E}^{(2)}$~\cite{SMAT}.
This means that once $C_2$ is known, the behavior for all $k$ follows.
In the large-$L_A$ limit,  $C_2$ converges exponentially fast to $2^{-L_A}$. Substituting into Eq.~\eqref{eq_avg_frame_1Tmeas} gives
\begin{equation}\label{eq_1T_frame_largeLA}
\mathbb{E}_{\mathcal{C} \sim \mathfrak{C}}\left[F_\mathcal{E}^{(k)} \right] \underset{L_A \gg 1}{=} 2^{-L_A}\Big(\frac{1}{2}+\frac{1}{2^k}\Big).
\end{equation}
Thus, a single $T$ gate is enough to induce a nontrivial $k$-dependence in the frame potential, but not sufficient to drive the projected ensemble toward a genuine state design, as after-all expected.

\begin{figure*}
    \centering
    \includegraphics[width=\linewidth]{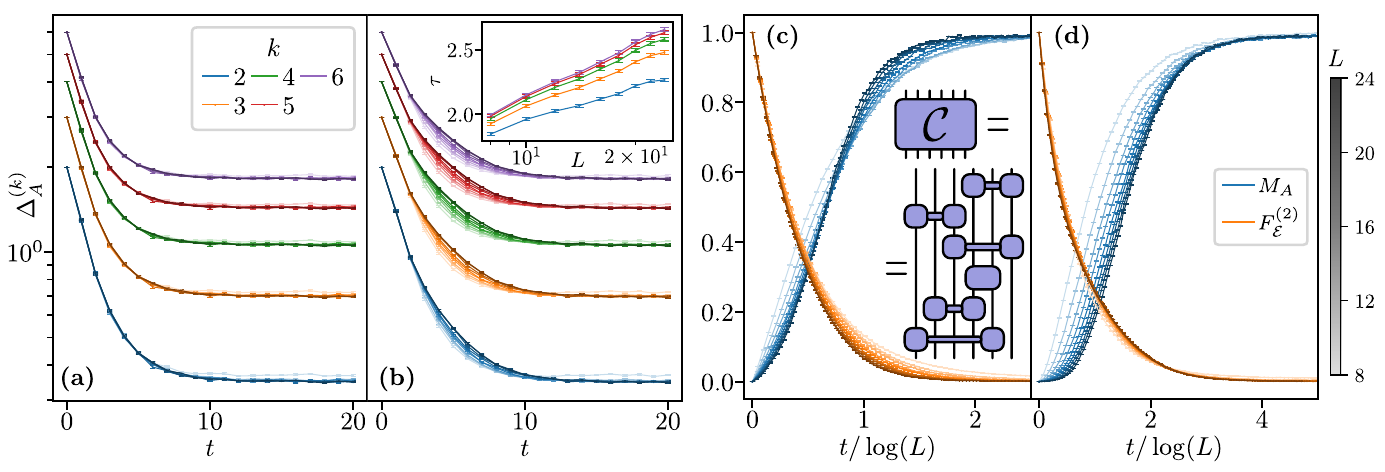}

    \vspace{-10pt}
    \caption{
    (a,b) Projected ensemble normalized distance to Haar $\Delta_A^{(k)}$, as a function of circuit depth $t$.  
    In the initial state $\ket{\psi_0^{\rm (left)}}$ of panel (a), magic is localized near subsystem $A$, whereas in the initial state $\ket{\psi_0^{\rm (right)}}$ of panel (b), it is concentrated on the opposite side. In both cases, $N_P = 1$ and $\theta = \pi/4$. In (b,inset) the saturation timescales are extracted by fitting $\Delta_A^{(k)} \sim \exp(-t/\tau)$. (c,d) Average Von Neumann magic in $A$ (blue) and second frame potential (orange), as a function of time $t$. For visualization purposes, both quantities are normalized to take values in $[0,1]$. A Clifford circuit with long-range connectivity $\alpha = 1$ (c) and $\alpha = 3$ (d) is used to produce the pre-measurement state (see inset in (c)). Different curves represent different values of $k$ and $L$, with fixed $L_A = 1$. Different choices of initial states and $L_A$ (with the same scaling of the distance between the magic injection and subsytem $A$) yield qualitatively identical behaviors. 
    }
    \label{fig_shallow}
\end{figure*}

To address the case of generic initial magic, we define single-qubit phase states $|\theta\rangle= P(\theta) H\ket{0} = (\ket{0} + e^{i\theta}\ket{1})/\sqrt{2}$, with $P(\theta) = e^{i\theta/2} e^{i\theta Z/2}$ the phase shift gate, 
and consider product states of the form
$|\psi_0(N_P,\theta)\rangle = |0\rangle^{\otimes (L-N_P)} \otimes |\theta\rangle^{\otimes N_P}$,
specified by the phase $\theta$ and the number of doped qubits ($P(\theta)$ gates) $N_P$.
The magic of these states is given by~\cite{Leone_2022,turkeshi2025pauli}
$M_k=N_P (1-k)^{-1} \log[(1+\cos^{2k}(\theta)+\sin^{2k}(\theta))/2]$. 
We compute $\Delta_A^{(k)}$ via exact simulations for various $k$, $N_P$, and $\theta$ (cf. Fig.~\ref{fig_deep}(a)). The key observation is that $\Delta_A^{(k)}$ decays exponentially with $N_P$, i.e. with the total magic of the initial state. Moreover, the decay rate is directly proportional to the stabilizer entropy density $M_k/N_P$ (while a decay with $M_2/N_P$ does not work as shown in panel (b)) for all $\theta$,  see Fig.~\ref{fig_deep}(b). 
Guided by the analytical results for a single $T$ gate and these numerical findings, we propose the following \emph{Magic-Induced Design Ansatz (MIDA)}, valid in the limit $L_A \gg 1$, 
\begin{equation}\label{eq_MIDA}
    \mathbb{E}_{\mathcal{C} \sim \mathfrak{C}}\left[\Delta_A^{(k)}(M_k)\right] = \mu(k,L_A) \exp[-\nu(k,L_A) M_k]\;.
\end{equation}
Here $M_k$ is the stabilizer entropy of the global initial state, $\mu=\Delta_A^{(k)}(M_k=0)$ is the stabilizer baseline [cf. Eq.~\eqref{eq_mu}], and $\nu(k,L_A)$ is determined from the single $T$-state case analytically up to the finite size factor $C_2$, cf.~\cite{SMAT},  
\begin{equation}\label{eq_nu}
    \nu(k, L_A) = (k-1) \frac{\log\left[ 1 + 2^{L_A}C_2\left( 1 + \tfrac{1}{\mu} \right) \left[ \left( \tfrac{1}{2} \right)^{k} - \tfrac{1}{2} \right]\right]}{\log\left[\frac{1}{2} + \left( \frac{1}{2} \right)^k \right]} 
     \, .
\end{equation}
In the limit $L_A\gg1$ with $L\gg L_A$, this reduces analytically to 
\begin{equation}\label{eq:resultFinal}
    \mathbb{E}_{\mathcal{C} \sim \mathfrak{C}} \left[ \Delta_A^{(k)} (M_k) \right] \underset{L_A \gg 1}{=} \mu \exp\left[ (1-k) M_k \right] \, ,
\end{equation}
with exponential convergence with $L_A$, cf. Eq.~\eqref{eq_1T_frame_largeLA} and~\cite{SMAT}.
Since $(1-k)$ is negative for $k>1$, this expression yields an exponential decay in $\Delta_A^{(k)}$ as $M_k$ increases.
As shown in Figs.~\ref{fig_deep}(c-f), MIDA shows a better agreement with the numerics with increasing $L_A$, reaching a strong matching already with $L_A = 6$.
See also Fig.~\ref{fig_chi2}, where we carried out additional robustness tests of the ansatz using a $\chi^2$-likelihood analysis, which quantitatively reveal an exponential approach to MIDA with $L$ and $L_A$.
We further tested MIDA using a different class of initial states
\begin{equation}
    |\psi_0(M)\rangle= |0\rangle^{\otimes L-4} \otimes |\phi_r(M)\rangle\;,
\end{equation}
where $|\phi_r(M)\rangle$ is a random state on $4$ qubits (size arbitrarily chosen) with fixed Von Neumann magic $M_1$ (see the protocol in~\cite{SMAT}). 
We used 40 different values of $M_1 \in [0,2]$ with associated $M_k(\ket{\psi_0(M)})$ computed numerically.
The results, shown in orange in Figs.~\ref{fig_deep}(c-f), confirm that the same exponential law holds, demonstrating the robustness of MIDA across different initial conditions.

To corroborate these numerics, we also study the frame potential deviation analytically.
In the supplement ~\cite{SMAT},  we present arguments that imply an upper bound on $\Delta_A^{(k)}(M_k)$ that is consistent with our ansatz \eqref{eq:resultFinal}:
\begin{equation}
    \label{eq: delta upper bound}
    \mathbb{E}_{\mathcal{C} \sim \mathfrak{C}} \left[ \Delta_A^{(k)}(M_k) \right] \leqslant 4 \frac{2^{k L_A}}{k!} 2^{-M_2(|\psi_0\rangle)}\;.
\end{equation}
To arrive at \eqref{eq: delta upper bound}, we employ the replica method, along with recent advances in the commutant theory of the Clifford group $\mathfrak{C}_N$~\cite{Gross2021,magni2025anticoncentration,bittel2025completetheorycliffordcommutant, bittel2025operationalinterpretationstabilizerentropy}.

\section{Shallow Clifford circuits and magic teleportation}
Our previous results provide a comprehensive picture of how much magic is required for a projected ensemble to approximate Haar random states under global Clifford evolution. This regime requires a depth $t=O(N)$, which is realistic for fault-tolerant devices but beyond the reach of present-day NISQ hardware. To connect with the latter, we next study shallow Clifford circuits. Fixing the initial magic content, we analyze how the projected ensemble approaches the asymptotic behavior as a function of circuit depth. A key finding is the emergence of \emph{magic teleportation}, a phenomenon directly analogous to the entanglement teleportation explored in~\cite{PhysRevLett.132.030401,McGinley2025MIEComplexity,Hoke2023}: the combination of unitary time evolution and measurements violates the typical light-cone locality and allows for transferring information via longer time-scales compared to simple unitary evolution. 

We study two distinct spatial configurations for the initial magic resource. In the first case,  
$|\psi_0^\mathrm{(left)}\rangle = \ket{0}^{\otimes L_A} \otimes |\theta\rangle^{\otimes N_P} \otimes |0\rangle^{\otimes L - N_P - L_A}$,  
the magic states are placed near subsystem $A$. In the second case,  
$|\psi_0^{\mathrm{(right)}}\rangle = |0\rangle^{\otimes L - N_P} \otimes |\theta\rangle^{\otimes N_P}$,  
the magic resources are initialized on qubits located far from $A$. We begin by analyzing a standard brickwork circuit, where each layer is given by 
\begin{equation}
    U_\mathrm{layer} = \left(\prod_{x=1}^{L/2-1} U_{2x,2x+1}\right)\left( \prod_{x=1}^{L/2} U_{2x-1,2x}\right)\;.
\end{equation}
We study how $\Delta_A^{(k)}(\ket{\psi_t}) $ evolves as a function of the circuit depth $t$. The results in Figs.~\ref{fig_shallow}(a,b) reveal a sharp contrast in timescales. When the magic in the initial state is adjacent to $A$, the discrepancy $\Delta_A^{(k)}$ decays rapidly and saturates after a constant depth, i.e. $O(1)$ in system size. Curves of $\Delta_A^{(k)}(t)$ for different $L$ collapse onto one another, showing that only a finite scrambling depth is required for the ensemble to reach its asymptotic randomness when the initial state is supported within or near $A$ (Fig.~\ref{fig_shallow}(a)). 
In contrast, when the magic is injected far away from $A$ (case b), the approach to saturation slows down with increasing $L$. By fitting $\Delta_A^{(k)} \sim e^{-t/\tau}$ (Fig.~\ref{fig_shallow}(b, inset)), we extract a convergence timescale $\tau$ that grows logarithmically with the distance of the magic from $A$, i.e. $\tau = O(\log L)$ for a system of length $L$. 
In the End Matter, we analyze an alternative protocol in which the system is initialized in $|\psi_0\rangle = |0\rangle^{\otimes L}$ and the last $N_P$ qubits are measured in $|\pm\theta\rangle=(|0\rangle\pm e^{i\theta}|1\rangle)/\sqrt{2}$, thereby injecting non-stabilizerness. Quite remarkably, analytical and numerical arguments confirm that this setup yields the same behavior as initializing the system in $|\psi_0^{\mathrm{(right)}}\rangle$ and subsequently performing projective measurements on $B$, see Fig. \ref{fig_rot_meas}.

Lastly, we investigate the long-range Clifford circuits of \cite{PhysRevLett.132.030401}, where we evolve under layers where $L \Delta t$ random qubit pairs $(x,y)$ are selected with probability $p(|x-y|)\propto |x-y|^{-\alpha}$ ($x\neq y$), each acted upon by a random two-qubit Clifford $U_{x,y}\sim\mathfrak{C}_2$, with $\Delta t$ the time step associated to each layer and periodic boundary conditions (see Fig.~\ref{fig_shallow}(c,inset)).
We start from the state $|\psi_0\rangle = \ket{0}^{\otimes(L/2 -2)} \otimes \ket{\phi_r(2)} \otimes \ket{0}^{\otimes(L/2-2)}$, meaning the magic is injected in the middle of the chain.
For strong long-range interactions ($\alpha\lesssim 2$), the projected ensemble converges to the stationary global Clifford value in constant depth, $O(1)$, while for suppressed long-range couplings ($\alpha\gtrsim 2$) the dynamics reduce to the local case with logarithmic saturation, $\tau = O(\log L)$, in agreement with the entanglement transition found in \cite{PhysRevLett.132.030401}. 
This is evident by analysing the finite-size shifts in Figs.~\ref{fig_shallow}(c,d).
While in Fig.~\ref{fig_shallow}(c), with $\alpha = 1$, the saturation times rescaled by $\log L$ shift towards the left, in Fig.~\ref{fig_shallow}(d), with $\alpha = 3$, the same rescaled saturation times converge with $L$.
Taken together, these results reflect a form of magic teleportation requirements. 
In local circuits, magic must propagate to generate a global approximate design~\cite{magni2025anticoncentration,grevink2025glueshortdepthdesignsunitary,Zhang2025DesignsMagicAugClifford} so that measurements effectively induce a state design on $A$. 
This mimics the case of random Haar circuits that was previously analyzed in~\cite{Lami2025}, with the important difference that it applies to doped Clifford circuits, of relevance in present architectures.

\section{Discussion}
We have characterized the emergence of randomness in projected ensembles generated by Clifford circuits doped with a finite amount of magic. Our analysis establishes a direct quantitative relation between the design order and the stabilizer entropy of the initial state: in the deep-circuit limit, the deviation from Haar randomness decays exponentially with $M_k$, interpolating smoothly between stabilizer states (1-designs) and ensembles with higher-order correlations. This behavior is captured by our MIDA formula, supported by both numerical data and rigorous bounds derived from Clifford commutant theory.
One can naturally extend our ideas by studying the projected ensemble randomness with respect to other kinds of quantum resources (for example, coherence \cite{liu2025coherenceinduceddeepthermalizationtransition}).
The existence of a correspondence to the MIDA for other quantum resources remains an open question for future work.

Going beyond global circuits, we analyzed shallow architectures and identified the role of magic teleportation in randomness generation. In local circuits, saturation times grow logarithmically with system size, reflecting the need for magic to propagate across the system before inducing design properties on $A$. By contrast, in long-range architectures, nonlocal gates allow magic to “jump” across the system, collapsing this light-cone and enabling randomness generation in constant depth. These findings directly connect the behavior of doped Clifford circuits to the phenomenology previously observed in Haar random circuits, but now in a setting directly relevant to near-term fault-tolerant devices.

\begin{acknowledgments}
\section{Acknowledgments} 
We thank A. De Luca, S. Ghosh,  A. Hamma, M. Heinrich, B. Magni, P. Sierant, T. Haug and E. Tirrito for discussions. 
XT acknowledges support from DFG under Germany's Excellence Strategy - Cluster of Excellence Matter and Light for Quantum Computing (ML4Q) EXC 2004/1 – 390534769, and DFG Collaborative Research Center (CRC) 183 Project No. 277101999 - project B01.
J.D.N., G.L., and H.L. are funded by the ERC Starting Grant 101042293 (HEPIQ) and the ANR-22-CPJ1-0021-01. MM acknowledges support from Trinity College, Cambridge.

\paragraph{Code and Data Availability.} The code and the data for our simulations will be publicly shared at publication. 

\paragraph{Author Contribution}
Hugo Lóio conducted all numerical simulations and most analytical calculations, assisted by Guglielmo Lami and Max McGinley. Lorenzo Leone designed the analytical argument for the matching with rotated measurements, and together with Guglielmo Lami, arrived at the upper bound for the randomness. Jacopo De Nardis and Xhek Turkeshi originally designed the work, supervised and, together with Hugo Lóio, wrote the manuscript.

\end{acknowledgments}

\bibliography{biblio}

\clearpage
\appendix 

\twocolumngrid
\begin{center}
    \textbf{\large End Matter}
\end{center}

\begin{figure}
    \centering
    \includegraphics[trim={12 15 13 13}, clip, width=\columnwidth]{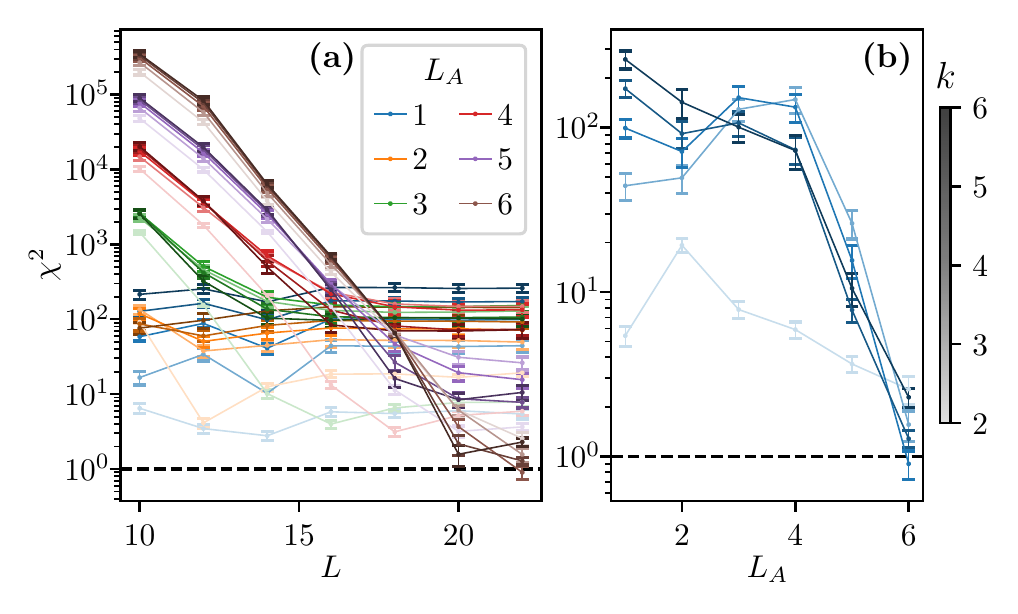}

    \vspace{-10pt}
    \caption{
    (a) Reduced chi-square as a function of $L$, for different $k$ and $L_A$.
    (b) Reduced chi-square as a function of $L_A$, for different $k$ and fixed $L = 22$.
    In dashed, $\chi^2 = 1$ signalling the expected value for perfectly compatible datasets.
    $N_0 = 80$ different initial states with $M \in  [0.2, 2.1]$ were used for the computation of $\chi^2$ (states with $M \ll 1$ removed due to disproportionately affecting $\chi^2$ with low statistical uncertainties, revealing mostly finite-size errors in $\mu$).
    }
    \label{fig_chi2}
\end{figure}

\noindent
\section{Statistical analysis of the goodness of MIDA}\label{sec_ansatz_dist} \\

\noindent
For fixed $L$, $L_A$, and $k$, we consider the set of asymptotic values of $\Delta_A^{(k)}$ at large time. We denote $\{y_i \pm \delta_i^y\}_{i=1}^{N_0}$ the numerical data and $\{\tilde{y}_i \pm \delta_i^{\tilde{y}} \}_{i=1}^{N_0}$ the prediction of MIDA. We consider different possible initial states, and $N_0$ is their total number. Note that there is some uncertainty in the MIDA due to the numerically estimated parameter $C_2$ in Eq.~\eqref{eq_nu} (although we ensured that $\delta_i^{\tilde{y}} \ll \delta_i^y$). To quantitatively determine how compatible the numerical data and MIDA predictions are, we utilize the reduced $\chi^2$ statistic, defined as
\begin{equation}\label{eq_chi2}
    \chi^2 = \frac{1}{N_0} \sum_i  \frac{(y_i - \tilde{y}_i)^2}{\delta_i^y + \delta_i^{\tilde{y} } }
    \, .
\end{equation}
The closer $\chi^2$ is to 1, the more compatible the datasets are. Our results are presented in Fig.~\ref{fig_chi2}, where we observe that for the smaller $L_A$ and $k > 2$, there is a convergence to $\chi^2 \gg 1$ with system size. 
However, for the larger values of $L_A$, $\chi^2$ decays exponentially with $L$. 
In Fig.~\ref{fig_chi2}(b), we fix the largest simulated system size $L$ and observe that there is approximately an exponential decay of $\chi^2$ with $L_A$ (statistical fluctuations are expected in $\chi^2$ due to the limited number of initial states). 
Already at $L_A = 6$, for most $k$, $\chi^2 \sim 1$, signaling near perfect compatibility between the ansatz and the numerical data. \\

\noindent
\section{Magic injection via rotated measurements} \\

\begin{figure}
    \centering
    \includegraphics[width=\columnwidth]{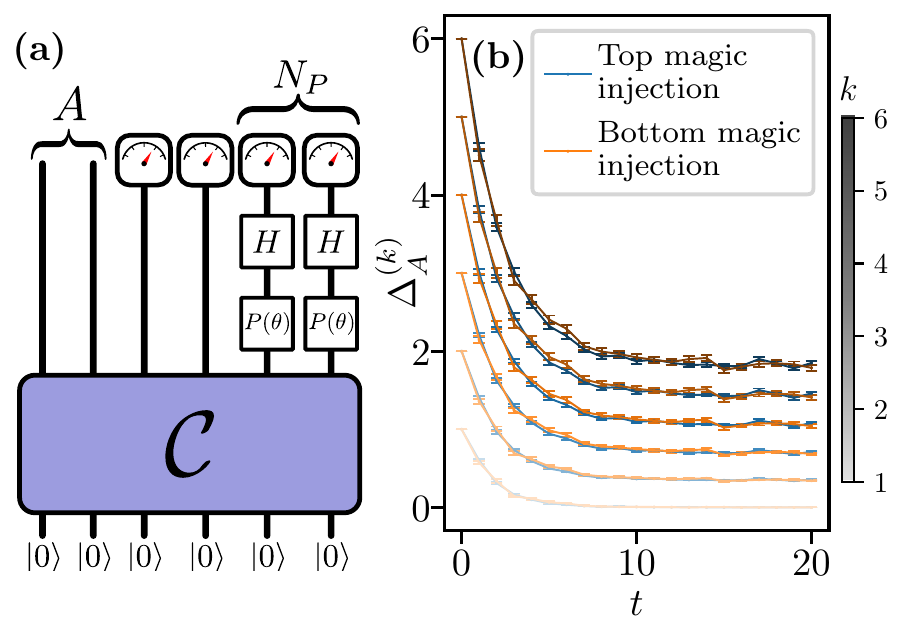}
    
    \vspace{-10pt}
    \caption{
    (a) Scheme for a modified version of the protocol in Fig. \ref{fig_protocol_scheme}. 
    In this case, the initial state is just a stabilizer product state such as $\ket{0}^{\otimes L}$.
    Instead, the magic is injected upon rotation of the measurement basis with $N_P$ phase shift gates $P(\theta)$ and Hadamard gates $H$.
    The measurements are performed in the $Z$ basis and the projected ensemble is obtained in subsystem $A$.
    (b) Time-dependent comparison for $\Delta_A^{(k)}$ between the protocol of (a) (in blue, i.e. magic injected at the top) and the protocol of Fig. \ref{fig_protocol_scheme}(a) with initial state $\ket{\psi_0^{\rm right}}$ (in orange, i.e. magic injected at the bottom), for different $k$.
    We fix $L_A = N_P = 1$, $\theta = \pi/4$, $L = 16$, although other parameter choices result in the same matching behaviour between the protocols.
    }
    \label{fig_rot_meas}
\end{figure}

\noindent

Here, we consider a modified protocol in which both the initial state and the circuit are free of magic resources, and magic is injected only just before the final measurements, see Fig.~\ref{fig_rot_meas}(a). Equivalently, the protocol can be interpreted as measuring a stabilizer state in a rotated magic basis. \\
Note that there is a matching of the local distribution of magic gates between the modified protocol of Fig.~\ref{fig_rot_meas}(a) and the standard protocol of Fig.~\ref{fig_protocol_scheme}(a) with the initial state $\ket{\psi_0^{\rm right}}$.
Surprisingly, Fig.~\ref{fig_rot_meas}(b) shows that this correspondence also reveals itself in a precise time-dependent matching of $\Delta_A^{(k)}$ between the protocols.
This leads us to claim that the randomness of the projected ensemble at any circuit depth depends on where the magic gates are locally applied, but not on the injection depth.
We support the statement with the following analytical argument, claiming that one can switch the Clifford circuit with the phase gates, with at most an $O(D)$ increase of circuit depth $D$.
A stricter proof of our data collapse could be the subject of future works.

\noindent
\textbf{Theorem.} Consider the state $\ket{\psi(N_P, D)}=\prod_{j=1}^{N_P}e^{i\sigma_j\theta_j}\mathcal{C}_D\ket{0}$, where $\sigma_j$ are local Pauli matrix operators acting on site $j$ (the complex exponential corresponding to $N_P$ local phase gates with corresponding phase $\theta_j$) and $\mathcal{C}_D$ is a Clifford operation of depth $D$. Then the following identity is true
\begin{equation}
    \ket{\psi(N_P,D)}=\mathcal{C}_{D(N_P)}\left[\otimes_{j=1}^{N_P'}\ket{\phi(\theta_j)}\otimes \ket{0}^{\otimes n-N_P'}\right]
\end{equation}
where $N_P'\le N_P$,  $\ket{\phi(\theta_j)}=\cos\frac{\theta_j}{2}\ket{0}+\sin\frac{\theta_j}{2}\ket{1}$, and $D(N_P) \leqslant 19D+9N_P$.

\textit{Proof.} 
We can write the state as
\begin{equation}
    \ket{\psi(N_P,D)}=\mathcal{C}_D \prod_{j=1}^{N_P}e^{i s_j\theta_j}\ket{0}^{\otimes n}
\end{equation}
where $s_j \in \mathcal{P}_L$ are obtained by the twirling of $\sigma$ via $\mathcal{C}_D$. 
Notice that all the $s_j$ commute with one another.
Now, for each $s_j$, two situations can happen: (i) either it acts trivially on $\ket{0}$, which would effectively reduce the number of non-Clifford gates to $N_P'\le N_P$, or (ii) it does not act trivially.
Let us assume that $N_P'$ $s_j$s act nontrivially. This necessarily means that there exists an element of the $\mathbb{Z}$ group generating $\ket{0}$ for which each $s_j$ individually anti-commutes. 
It follows that there exists a Clifford $\mathcal{C}'$ such that
\begin{equation}
    \mathcal{C}^{\prime\dagger}s_{j}\mathcal{C}^{\prime}=X_j \quad , \quad
    \mathcal{C}^{\prime}\ket{0}^{\otimes n}=\ket{0}^{\otimes n}
\end{equation}
where $X_j$ is the $X$ Pauli matrix acting on the site $j$.
The existence of the above Clifford follows from the fact that all $s_j$ commute with each other and all of them anti-commute with the $\mathbb{Z}$ group.
This implies we can write
\begin{equation}
    \ket{\psi(N_P,D)}=\mathcal{C}_D\mathcal{C}'\prod_{j=1}^{N_P'}e^{iX_j\theta_j}\ket{0}^{\otimes n}
\end{equation}
It just suffices to bound the depth of $\mathcal{C}'$.
We notice that each $s_j$ has been constructed by a single qubit Pauli matrix twirled by a depth $D$ nearest-neighbor Clifford $\mathcal{C}_D$, meaning that it can have at most support (i.e.\ being non-identity) on $2D$ many sites. 
The Clifford $\mathcal{C}'$ is defined collectively on all the $s_j$, which altogether are different from the identity on a support $2D+N_P'$. Hence, the Clifford $\mathcal{C}'$ can be defined only on $2D+N_P$ qubits.
To conclude the proof, we note that any Clifford circuit on $N$ qubits can be implemented with depth at most $9 N$ in a nearest-neighbor architecture~\cite{9435351}. Hence, the Clifford $\mathcal{C}'$ has depth at most $9(2D+N_P)$, and the overall Clifford has depth $19D+9N_P$, which proves the statement.

\onecolumngrid
\newpage

\begin{center}
\textbf{\large Supplemental Material}
\end{center}
In this Supplemental Material, we:
\begin{itemize}
    \item Describe the numerical protocol established for generating random states with fixed finite magic.
    \item Derive the projected ensemble frame potential of a random stabilizer state.
    \item Extend the previous point with a single $T$-gate doping.
    \item Derive a bound on the asymptotic Haar distance for a generic initial state with finite magic. 
\end{itemize}

\begin{figure}[h!]
    \centering
    \includegraphics[width=0.6\columnwidth]{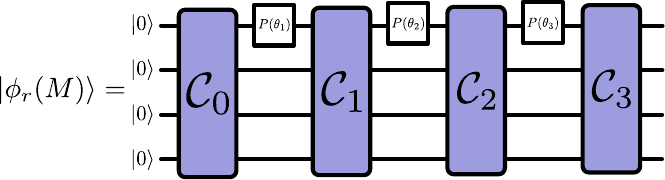}
    \caption{
    A Clifford circuit with global Clifford gates $\mathcal{C}_i$ doped with phase gates $P(\theta_i)$.
    An appropriate choice of the phases $\theta_i$ allows us to fix the Von Neumann SE of the state after the protocol.
    }
    \label{fig_random_magic_state_protocol}
\end{figure}

\section{Sampling random states with fixed stabilizer entropy}\label{sec_random_state_protocol}

We design a protocol for producing random states with a target Von Neumann stabilizer entropy $M = \lim_{k\rightarrow 1} M_k$,  depicted in Fig. \ref{fig_random_magic_state_protocol}. 
We set the target magic with a tolerance $\epsilon$.
The protocol starts with a random stabilizer state (global random Clifford $\mathcal{C}_0$ applied on a product state) such that $M^{(0)} = 0$.
Each step $i$ in the algorithm consists of
\begin{enumerate}
    \item Apply the random global Clifford gate $\mathcal{C}_i$.
    \item Compute $\Delta M = M - M^{(i-1)}$.
    \item If $\Delta M < \epsilon$, return the final state.
    \item If $\Delta M \geqslant \log(2)/2$ choose $\theta_i = \pi/4$. 
    Else, find $\theta_i$ s.t. $\Delta M = -(\sin^2\theta_i \log\sin\theta_i + \cos\theta_i \log\cos^2\theta_i$.
    (This step assumes heuristically that each phase gate can produce an increment of magic bounded by the magic of $P(\theta_i)H\ket{0}$.)
    \item Compute the magic $M^{(i)}$ of the state.
    If $M^{(i)} > M$, apply the hermitian conjugate of the previous phase gate and set $M^{(i)} = M^{(i-1)}$.
    (We have noticed that sometimes the magic overshoots the target, possibly because either the assumed bound in the previous step occasionally fails or due to the numerical precision of the root finding method for $\theta_i$.)
\end{enumerate}
We have observed that $M^{(i)}$ will monotonically approach $M$ as long as $M$ is smaller than the typical magic of a random Haar state of the same size.
This protocol will produce random states with the same target magic $M$, but it will not fix the stabilizer Rényi entropies $M_k \ , \ k>1$.
The distribution of these states in the manifold of possible target states is unknown.

\section{Projected ensemble of a stabilizer state}\label{sec_stab_meas}

Consider an $L$-qubit stabilizer state $\ket{\psi}$ with stabilizer group $S(\ket{\psi}) = \langle s_1, s_2, \dots, s_L \rangle$, where $s_i \in \{\pm 1, \pm i\} \times \mathcal{P}_L$ are the generators.
If we measure $\ket{\psi}$ on site $i$ in the $Z$-basis, there are two possible cases to consider:
\begin{enumerate}
    \item $Z_i$ commutes with all the generators, $[s_j, Z_i] = 0 \, , \forall j \in \{1,\dots, L\}$.
    In this case $Z_i \ket{\psi}$ is stabilized by the same generators as $\ket{\psi}$, so we can conclude $Z_i \ket{\psi}  = \pm \ket{\psi}$.
    Measuring does not alter the state and $\pm Z_i$ stabilizes the state $\ket{\psi}$ (the sign agrees with the measurement outcome).
    \item There is one generator $s_j$ which anti-commutes with $Z_i$, $\{s_j , Z_i\} = 0$.
    The generators can be chosen such that only a single generator anti-commutes with $Z_i$, since if $s_l$ also anti-commutes with $Z_i$, then we can update the generator as $s_l \rightarrow s_js_l$ so that it commutes instead. \\
    In this case, the measurement outcome probabilities are given by
    \begin{equation}
    p(+1) = \bra{\psi} \frac{1 + Z_i}{2} \ket{\psi} = 
    \bra{\psi} s_j^\dagger \frac{1 + Z_i}{2} s_j \ket{\psi} =
    \bra{\psi} \frac{1 - Z_i}{2} \ket{\psi} =
    p(-1)
    \ , 
    \end{equation} 
    so we can conclude that $p(+1) = p(-1) =1/2$ and the post-measurement state is stabilized by $\langle s_1, \dots, s_{j-1}, \pm Z_i, s_{j+1}, \dots, s_L\rangle$ \cite{Nielsen_Chuang_2010}.
\end{enumerate}

Note that in both cases, the post-measurement state is stabilized by $\pm Z_i$, and the remaining generators, which are also generators of the pre-measurement state, can only be $\mathds{1}$ or $Z$ at site $i$.

If we iteratively measure many sites, the surviving generators of the pre-measurement state will always be diagonal in the measured subspace.
If we measure out a subsystem as in Fig. \ref{fig_protocol_scheme}, we get the stabilizer state generated by
\begin{equation}
    S(\ket{\psi(z_B)}) = \langle s_1, \dots, s_{L_A}, \alpha'_1(z_B) Z_{L_A + 1}, \dots, \alpha'_{L_B}(z_B) Z_{L} \rangle
    \ , \ \alpha'_i(z_B) \in \{-1,1\} \ \ ,
\end{equation}
where the generators $s_i$ are diagonal in $B$.
We can then get rid of the non-trivial diagonal elements of $s_i$ in subspace $L_B$ by composing them with the generators $\pm Z_j$. 
Then the state can be trivially reduced to subspace $A$, 
\begin{equation}
    S(\ket{\psi_A(z_B)}) = \langle \alpha_1(z_B) s^A_1, \dots, \alpha_{L_A}(z_B) s^A_{L_A} \rangle \ , \ \alpha_i(z_B) \in \{-1,1\} \ ,
\end{equation}
where $s^A_i$ are the substrings in subsystem $A$ of the generators $s_i$ of the pre-measurement state.
Consequently, generators of the stabilizer group for different measurement outcomes differ only by signs.
It then follows that 
\begin{equation}\label{eq_overlap_post_meas_P0}
  \langle \psi_A(z_B') | \psi_A(z_B) \rangle
  = \alpha_i(z_B) \langle \psi_A(z_B') |  s^A_i | \psi_A(z_B) \rangle 
  =  \alpha_i(z_B) \alpha_i(z_B') \langle \psi_A(z_B') | \psi_A(z_B) \rangle 
  \, ,
\end{equation}
meaning that the overlaps between post-measurement states are only not zero when $\alpha_i(z_B) = \alpha_i(z_B') \,, \, \forall \, i \in \{1, L_A\}$. 
In that case, the stabilizer groups are identical, so they correspond to the same state, and the overlap is one.
Consequently, the overlaps of two post-measurement states have just two possible outcomes $\left| \langle \psi_A(z_B) | \psi_A(z_B') \rangle \right| \in \{0,1\}$.

\section{Asymptotic frame potential in a single $T$-gate doped state}\label{sec_1T_details}

The initial state with $N_P = 1$ and $\theta = \pi/4$ (a single $T$-gate) is 
$ \ket{\psi_0} = \left( \ket{0} + e^{i\frac{\pi}{4}} X_L\ket{0}\right) / \sqrt{2}$, 
meaning the time-evolved state at large circuit depth is given by
\begin{equation}
    \ket{\psi} = \frac{1}{\sqrt{2}} \left( \mathcal{C}\ket{0} + e^{i\frac{\pi}{4}} \mathcal{C}X_L\ket{0}\right) 
    = \frac{1}{\sqrt{2}} \left( \ket{\psi^1} + e^{i\frac{\pi}{4}} \ket{\psi^2} \right) 
    \ ,
\end{equation}
where $\ket{\psi^1}, \ket{\psi^2}$ are properly normalized stabilizer states.
The stabilizer group for the $\ket{0}$ state is generated by $\langle Z_1, \dots, Z_L \rangle$, while the one for the $X_L\ket{0}$ state is generated by $\langle Z_1, \dots, Z_{L-1}, - Z_L \rangle$.
The generators evolve under the action of the Clifford circuit as $Z_i \rightarrow \mathcal{C} Z_i \mathcal{C}^\dagger$.
Therefore, the stabilizer group generators of the two stabilizers $\ket{\psi^1}$ and $\ket{\psi^2}$  can be chosen to be the same, except for a single generator with opposite sign. 
Composing generators to form a new set can propagate the sign of the last generator but it can also help understand the measurement procedure.
A more useful set of generators is 
\begin{equation}\label{eq_pre_meas_generators_xhek_NT1}
    S(\ket{\psi^\nu}) = \left\langle 
    \alpha_1^\nu s_1, \dots, \alpha_{L_A}^\nu s_{L_A}, 
    \alpha^\nu_{r_1} s_{r_1}, \dots, \alpha^\nu_{r_{N_r}} s_{r_{N_r}}, 
    \alpha^\nu_{d_1} s_{d_1}, \dots, \alpha^\nu_{d_{N_d}} s_{d_{N_d}} 
    \right\rangle
    \ , 
\end{equation}
where $\alpha^\nu_i \in \{-1, 1\}$ are signs, such that all the Pauli strings are unsigned.
The sets $\{r_i\}_i$ and $\{d_i\}_i$, with sizes $N_r$ and $N_d$ respectively, are two disjoint sets of site indices comprising subsystem $B$.
The sites $d_i$ correspond to the ones where measurements with deterministic outcomes will occur, whilst in sites $r_i$ they will occur with fully random outcomes.
Note that the sets $\{r_i\}_i$ and $\{d_i\}_i$ depend on the chosen order of measurements but are not random.
After a random measurement in a site $r_i$ the generator $\alpha_{r_i}^\nu s_{r_i}$ is replaced by $\pm Z_{r_i}$.
Once we reach a site $d_i$ we know that the generators $s_{d_i}$ can be composed with the rest to become $Z_{d_i}$. 
The Pauli strings $s_i\ , \ i \in \{1,\dots, L_A\}$, are the ones which will remain after the measurement procedure, meaning that they are diagonal in subspace $B$.

Let us now understand what happens to the post-measurement state of a specific possible outcome $z_B$.
If $\exists i \, : \, \alpha_{d_i}^1 \neq \alpha_{d_i}^2$, then $\ket{\psi^1}$ and $\ket{\psi^2}$ do not have compatible measurement outcomes and the post-measurement state is always a pure stabilizer, i.e. we retrieve the no-magic case discussed before, $| \langle \psi_A(z_B) | \psi_A(z_B') \rangle| \in \{0,1\}$.
In this case, the Born rule probability in each $r_i$ site is $1/2$ and the Born rule probability in each site $d_i$ is $1$ except when one of the stabilizers is projected to 0, in which case it is $1/2$. 
Therefore, $p(z_B) = 2^{-N_r-1}$, meaning there are $2^{N_r + 1}$ equally likely measurement outcomes.

Let us assume then that the opposite occurs, i.e. $\alpha^1_{d_i} = \alpha^2_{d_i} \ \forall \ i \in \{1, N_d\}$, so the deterministic measurements do not change the state (they always have Born rule probability $1$).
The post-measurement stabilizers are generated by
\begin{equation}\label{eq_xhek_1T_post_meas_stab_groups}
    S(\ket{\psi^\nu_A(z_B)}) = \left\langle 
    \alpha_1^\nu \beta_1(z_B) s_1^A, \dots, \alpha_{L_A}^\nu \beta_{L_A}(z_B) s^A_{L_A} 
    \right\rangle
    \ , 
\end{equation}
where the previous Pauli strings generators have been reduced to $A$, at the cost of an extra sign $\beta_i(z_B)$, which depends on the measurement outcome of the random measurements and on the content of $s_i$ in $B$ (which are the same for both $\ket{\psi^1}$ and $\ket{\psi^2}$).

If we consider the case where $\alpha_i^1 = \alpha_i^2 \ \forall \ i \in \{1,L_A\}$ the post-measurement stabilizers are the same up to a phase which can only be $\pm 1$ (since those are the eigenvalues of $X_L$).
The Born rule probabilities of sites $r_i$ are $1/2$, except for the site where the post-measurement stabilizers become proportional which gives Born probability $(1 \pm \sqrt{2}/2)/2$, depending on the relative phase.
There are then two possible total Born probabilities $p(z_B) = (1 \pm \sqrt{2}/2)2^{-N_r}$.

The last possibility is when $\ket{\psi_A^1(z_B)}$ and $\ket{\psi_A^2(z_B)}$ have different signs in the generators meaning they are orthogonal stabilizers and the post-measurement state becomes
\begin{equation}
    \ket{\psi_A(z_B)} = \frac{1}{\sqrt{2}} \left(\ket{\psi_A^1(z_B)} + e^{i\pi/4}\ket{\psi_A^2(z_B)} \right) \ .
\end{equation}
In this case, the Born rule probabilities do not depend on the measurement outcomes and are given by $p(z_B) = 2^{-N_r}$.

Considering distinct measurement outcomes, if $\beta_i(z_B') = \beta_i(z_B) \ \forall i$, then $|\langle{\psi_A^\nu(z_B)} | \psi_A^\nu(z_B')\rangle | = 1$ , meaning $| \langle \psi_A(z_B) | \psi_A(z_B') \rangle| \in \{0,1\}$.
Else if $\alpha^1_i\beta_i(z_B') = \alpha^2_i\beta_i(z_B) \ \forall i$, then $| \langle \psi_A(z_B) | \psi_A(z_B') \rangle| = \sqrt{2}/2$.
Otherwise, if none of the previous conditions holds, all the stabilizers involved in the overlap are orthogonal and the overlap is zero.

\begin{figure}
    \centering
    \includegraphics[width=0.8\columnwidth]{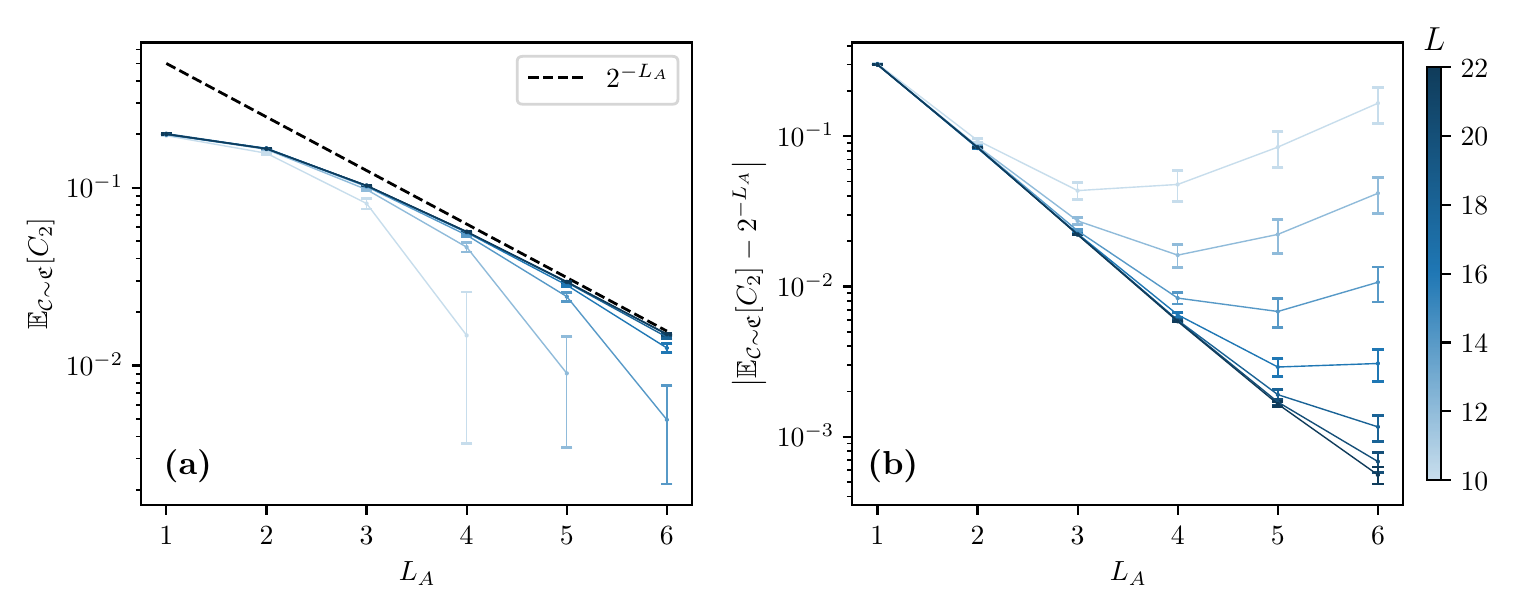}
    \caption{
    For an initial product phase state with $N_P = 1$ and $\theta = \pi/4$, $\mathbb{E}_{\mathcal{C} \sim \mathfrak{C}} [C_2]$ is estimated numerically by inverting \eqref{eq_avg_frame_1Tmeas}, as a function of $L_A$, for different system sizes $L$.
    (a) $\mathbb{E}_{\mathcal{C} \sim \mathfrak{C}}[C_2]$ as function of $L_A$, approaching $2^{-L_A}$ in dashed.
    (b) Convergence of $\mathbb{E}_{\mathcal{C} \sim \mathfrak{C}}[C_2]$ to $2^{-L_A}$ as a function of $L_A$.
    }
    \label{fig_c2}
\end{figure}

Therefore, the possible overlaps
which we obtain in the single $T$-gate case are $|\langle \psi_A(z_B) | \psi_A(z_B') \rangle| \in \{0,1 ,\sqrt{2}/2\}$. 
Consequently, the frame potential given by \eqref{eq_frame_potential} can then be expanded as
\begin{equation}\label{eq_frame_1T_C12}
    F_\mathcal{E}^{(k)} = C_1 + C_2 \frac{1}{2^k}  \ , 
\end{equation}
where $C_i$ are constants to be determined, meaning if we know the frame potential for $k \in \{1,2\}$, then we can determine it for larger $k$.
As explained in the main text, the first moment of the frame potential is always asymptotically known \eqref{eq_F1}, resulting (in the thermodynamic limit) in $\mathbb{E}_{\mathcal{C} \sim \mathfrak{C}}[C_1] + \mathbb{E}_{\mathcal{C} \sim \mathfrak{C}}[C_2]/2 = 2^{-L_A}$, which we can substitute into \eqref{eq_frame_1T_C12} to obtain \eqref{eq_avg_frame_1Tmeas}.
The value of $\mathbb{E}_{\mathcal{C} \sim \mathfrak{C}}[C_2]$ can be estimated numerically, as seen in Fig. \ref{fig_c2}, which seems to approach $2^{-L_A}$ exponentially fast in $L_A$.
Note that in the main text, we slightly change the notation as $\mathbb{E}_{\mathcal{C} \sim \mathfrak{C}}[C_2] \rightarrow C_2$ (the average becomes implicit).

Now that we have a good understanding of the measurement procedure, let us try to compute the overlap distributions of states in the projected ensemble. 
We will first attempt to compute the probabilities of each of the following possible cases (which depend only on the choice of the global Clifford gate):
\begin{itemize}
    \item \textbf{(i)} \textit{The measurements kill one of the stabilizers in $\ket{\psi}$.}
    
    \item \textbf{(ii)} \textit{The measurements merge the stabilizers in $\ket{\psi}$.}
    
    \item \textbf{(iii)} \textit{The stabilizers remain orthogonal after the measurements.}
\end{itemize}

It is first useful to compute the probability of finding a deterministic measurement at a given site $i$ (assuming all sites larger than $i$ have already been measured),
\begin{equation}
    P(i \in \{d_j\}_j ) =  2\frac{\#_S(i-1)}{\#_S(i)} = \frac{1}{2^i+1} \ ,
\end{equation}
where $\#_S(n) = 2^n\prod_{i=1}^n (2^i+1)$ are the number of $n$-qubit stabilizer states.
Then the probability of getting no deterministic measurements is
\begin{equation}
    P(N_d = 0) = \prod_{i = L_A +1}^L \left(1 - \frac{1}{2^i+1} \right) = \prod_{i = L_A +1}^L \frac{1}{1 + \frac{1}{2^i}} \underset{L_A \gg 1}{\simeq} 1 \ .
\end{equation}
Let us focus on the large $L_A$ limit, where there are no deterministic measurements and (i) does not occur.
Since we act on the initial state with a global random Clifford $C$, the signs $\alpha_i^\nu$ are fully random.
Moreover, the probability of finding (ii), which corresponds to the case $\alpha_i^1 = \alpha_i^2 \ \forall \ i \in \{1, \dots, L_A\}$ is also exponentially suppressed at large $L_A$.
We are then left with (iii), with Born rule probabilities $p(z_B) = 2^{L_A-L}$, and need only compute the probability distribution of each overlap,
\begin{equation}
\begin{aligned}
    & P\left(| \langle \psi_A(z_B) | \psi_A(z_B') \rangle| = 1\right) = 
    \frac{1}{2} P\left(\beta_i(z_B') = \beta_i(z_B) \ \forall \ i\right) = 2^{-L_A-1} \\
    & P\left(| \langle \psi_A(z_B) | \psi_A(z_B') \rangle| = \sqrt{2}/2 \right) = 
    P\left(\alpha_i^1 \beta_i(z_B') = \alpha_i^2 \beta_i(z_B) \ \forall \ i\right) = 2^{-L_A} \\
    & P\left(| \langle \psi_A(z_B) | \psi_A(z_B') \rangle| = 0 \right) =  1 - 3 \left( 2^{-L_A-1} \right)
\end{aligned}
\, .
\end{equation}
Note that the factor of $1/2$ in the first term comes from the fact that the condition $\beta_i(z_B') = \beta_i(z_B) \ \forall \ i$ ensures that $\ket{\psi_A^\nu(z_B)} = \pm \ket{\psi_A^\nu(z_B')}$, but it is only when $\ket{\psi_A^\nu(z_B)} = + \ket{\psi_A^\nu(z_B')}$ that the  $| \langle \psi_A(z_B) | \psi_A(z_B') \rangle| = 1$.
After applying the overlap distributions to \eqref{eq_frame_potential}, the frame potential at large $L_A$ gives \eqref{eq_1T_frame_largeLA} which agrees with the analytical prediction at $k=1$ and with the numerical prediction in Fig. \ref{fig_c2}, i.e. $\mathbb{E}_{\mathcal{C} \sim \mathfrak{C}}[C_2] \underset{L_A \gg 1}{\rightarrow} 2^{-L_A}$ .

\section{Upper bound on distance with generic initial magic states}
In this section, we present analytical arguments that substantiate the upper bound \eqref{eq: delta upper bound} presented in the main text. For this purpose, let us define the usual twirling operators for the Haar and Clifford groups as
\begin{equation}
\begin{aligned}
\Phi^{(m)}_{\Haar}(O) &= \mathbb{E}_{U\sim\Haar}[U^{\otimes m} O (U^{\dagger})^{\otimes m}] \\
\Phi^{(m)}_{\Cl}(O) &= \mathbb{E}_{U\sim\Cl}[U^{\otimes m} O (U^{\dagger})^{\otimes m}] \, .
\end{aligned}
\end{equation}
It is possible to compute the twirling operators above by the commutant theory over the unitary group and the Clifford group respectively. In particular, the commutant of the unitary group is spanned by permutation operators $\sigma \in S_m$, and consequently, the twirling reads
\begin{equation}
\Phi^{(m)}_{\Haar}(O) = \sum_{\sigma \in S_m} c^{\Haar}_{\sigma}(O) \sigma \, ,  \qquad  c^{\Haar}_{\sigma}(O) = \sum_{\sigma' \in S_m} \Wg^{\Haar}_{\sigma' \sigma} \tr[\sigma' O]
\end{equation}
where the matrix $\Wg^{\Haar}$ is given by the inverse of the Gram-Schmidt matrix $G_{\sigma'\sigma}=\tr(\sigma\sigma')$ and its components are called \textit{Weingarten functions}.
A similar expression can also be written for the Clifford group. The commutant of the Clifford group is spanned by the set of Pauli monomials $\Omega$~\cite{bittel2025completetheorycliffordcommutant,bittel2025operationalinterpretationstabilizerentropy}. Following the notation of Refs.~\cite{bittel2025completetheorycliffordcommutant,bittel2025operationalinterpretationstabilizerentropy}, we denote the set of these Pauli monomials as $ \mathcal{P} = \{ \Omega \}$. The twirling operators for the Clifford group can be expressed as: 
\begin{equation}
\Phi^{(m)}_{\Cl}(O) = \sum_{\Omega \in \mathcal{P}} c^{\Cl}_{\Omega}(O) \Omega \, , \qquad  c^{\Cl}_{\Omega}(O) = \sum_{\Omega' \in \mathcal{P}} \Wg^{\Cl}_{\Omega' \Omega} \tr[(\Omega')^{\dagger} O] \, .
\end{equation}
where, similarly to the case of the unitary group, the matrix $\Wg^{\Cl}$ is obtained by the inverse of the Gram-Schmidt matrix $G_{\Omega\Omega'}=\tr(\Omega\Omega')$ and its components are called \textit{Clifford-Weingarten functions}. The number of Pauli monomials can be bounded as (see Lemma 17~\cite{bittel2025operationalinterpretationstabilizerentropy})
\begin{equation}
|\mathcal{P}| = \prod_{j=0}^{m-2} (2^j + 1) \leq 2^{\frac{m^2}{2} - \frac{3}{2} m + 6} \, . 
\end{equation}
Now we have all the machinery to prove Eq.~(10). In our setup (see Fig. \ref{fig_protocol_scheme}), the generalized frame potential (averaged over the global Clifford) reads
\begin{equation}
F^{(k,n)}_{\Cl} = \tr[ Q^{(k,n)} \Phi^{(m)}_{\Cl}\big( \rho_0^{\otimes m} \big) ] \, .
\end{equation}
where 
\begin{equation}
Q^{(k,n)} = \sum_{z,z' \in \{0,1\}^{L_B}} (|z\rangle\langle z| \otimes |z'\rangle\langle z'|)^{\otimes n} (|z\rangle\langle z'| \otimes |z'\rangle\langle z|)^{\otimes k}
\end{equation}
and $m=2(n+k)$ is the total number of replicas. The conventional frame potential is recovered in the limit $n \to 1-k$ ($m \to 2$). We compare $F^{(k,n)}_{\Cl}$ with the corresponding value for Haar, namely
\begin{equation}
F^{(k,n)}_{\Haar} = \tr[ Q^{(k,n)} \Phi^{(m)}_{\Haar}\big(\rho_0^{\otimes m} \big) ] \, .
\end{equation}
The deviation is 
\begin{equation}
\left|F^{(k,n)}_{\Cl} - F^{(k,n)}_{\Haar}\right|= \left| \tr[ Q^{(k,n)} \bigg( \Phi^{(m)}_{\Cl}\big(\rho_0^{\otimes m}\big)  - \Phi^{(m)}_{\Haar}\big(\rho_0^{\otimes m} \big) \bigg)] \right|
\end{equation}
which, by standard matrix inequalities, can be bounded as 
\begin{equation}
\left|F^{(k,n)}_{\Cl} - F^{(k,n)}_{\Haar}\right| \leq \left\|Q^{(k,n)}\right\|_{\infty} \left\| \Phi^{(m)}_{\Cl}\big(\rho_0^{\otimes m} \big)  - \Phi^{(m)}_{\Haar}\big(\rho_0^{\otimes m} \big) ] \right\|_{1} \, .
\end{equation}
Now notice that 
\begin{equation}
Q^{(k,n)} = \sum_{Z=(z,z')} (|Z\rangle\langle Z|)^{\otimes n} (|Z\rangle\langle Z|)^{\otimes k} \text{SWAP} = \sum_{Z} (|Z\rangle\langle Z|)^{\otimes (n+k)} \text{SWAP}
\end{equation}
where $\text{SWAP}$ is inserted to swap $z'$ and $z$ in the last $2k$ replicas. Using the unitary invariance of the Schatten norms, we find 
\begin{equation}
\left\|Q^{(k,n)}\right\|_{\infty} = \left\|\sum_{Z} (|Z\rangle\langle Z|)^{\otimes (n+k)}\right\|_{\infty} = 1 \, .
\end{equation}
Now we have to deal only with the term $\left\| \Phi^{(m)}_{\Cl}\big(\rho_0^{\otimes m} \big)  - \Phi^{(m)}_{\Haar}\big(\rho_0^{\otimes m} \big)  \right\|_{1}$. Bounds for this term can be found in Theorem 14 of~\cite{bittel2025operationalinterpretationstabilizerentropy}. In particular, for all $k \geq 4$, we have
\begin{equation}
2^{-2 M_{3} (|\psi_0 \rangle)} - \frac{1}{2^L} - \frac{16}{2^{2L}} \leq \left\| \Phi^{(m)}_{\Cl}\big(\rho_0^{\otimes m} \big)  - \Phi^{(m)}_{\Haar}\big(\rho_0^{\otimes m} \big)  \right\|_{1}  \leq 2^{\frac{m^2}{2}} 2^{-M_{2} (|\psi_0 \rangle)} + 2^{2m^2 - L}  \, .
\end{equation}
We are interested in the normalized deviation of the $k$-th frame potential, namely
\begin{equation}
\Delta^{(k)}_A = \frac{\left|F^{(k)}_{\Cl} - F^{(k)}_{\Haar}\right|}{F^{(k)}_{\Haar}} \simeq \frac{2^{k L_A}}{k!} \lim_{n \to 1-k} \left|F^{(k,n)}_{\Cl} - F^{(k,n)}_{\Haar}\right|
\end{equation}
Using the above inequalities (with $m \to 2$), and neglecting all terms that are exponentially vanishing for large $N$, we arrive at
\begin{equation}
\Delta^{(k)}_A  \leq 4 \frac{2^{k L_A}}{k!} 2^{-M_{2} (|\psi_0 \rangle)} \equiv \overline{\Delta_A^{(k)}}\, .
\end{equation}
If we define the stabilizer purities as 
\begin{equation}
\spur_{2 k}(|\psi \rangle) = e^{\left[ (1- k) M_{k} (|\psi \rangle) \right]} = 2^{-L} \sum_{s \in \mathcal{P}_L} \langle \psi | s | \psi \rangle^{2 k} \, ,
\end{equation}
where $s$ are the Pauli strings and $\mathcal{P}_L=\lbrace I,X,Y,Z\rbrace^{\otimes L}$, we find
\begin{equation}
\frac{2^{k L_A}}{k!} \spur_{4}(|\psi_0 \rangle) \leq \overline{\Delta_A^{(k)}}\, .
\end{equation}
Since $\spur_{k}(|\psi_0 \rangle)  \leq \spur_4(|\psi_0 \rangle)$, then $\exp \left[ (1-k) M_{k} (|\psi \rangle) \right] \leq \exp \left[ - M_{2} (|\psi \rangle) \right] $ for $k \geq 4$, which makes this inequality compatible with MIDA that reads, for $2^{L_A} \gg k$,
\begin{equation}
\Delta_A^{(k)} = \frac{2^{(k-1) L_A}}{k!}e^{ (1-k) M_{k} (|\psi \rangle)} \leq \overline{\Delta_A^{(k)}}\, .
\end{equation}
Although the inequalities derived above hold rigorously for all  $n, k \geq 1$, the analytic continuation to the limit $n \rightarrow 1 - k$ is not guaranteed to preserve the validity of the bound. Nevertheless, the bound is confirmed by our numerics. Moreover, for sufficiently large $A$, the measurement outcome probabilities $p(z_B)$ will be close to uniform as a result of anticoncentration; thus, the dependence of all relevant quantities on $n$ will be parametrically weak. It is therefore unlikely that this non-rigorous step is crucial.

\end{document}